# Mechanics of an asymmetric hard-soft lamellar nanomaterial

**Weichao Shi[†], Glenn H. Fredrickson[†,‡,*], Edward J. Kramer[†,‡,§,○]**

[†]*Materials Research Laboratory,* [‡]*Department of Chemical Engineering,* [§]*Department of Materials, University of California at Santa Barbara, California, United States, 93106*

**Christos Ntaras[#] and Apostolos Avgeropoulos[#,*]**

[#]*Department of Materials Science and Engineering, University of Ioannina, University Campus, Ioannina, Greece 45110*

**Quentin Demassieux[‖] and Costantino Creton[‖]**

[‖]*Laboratory of Soft Matter Science and Engineering, ESPCI Paristech-CNRS-UPMC, 10 rue Vauquelin, 75005 Paris, France*

[○]Edward J. Kramer passed away on Dec. 27[th], 2014.

# Abstract

Nano-layered lamellae are common structures in nanoscience and nanotechnology, but most are nearly symmetric in volume fraction. We report the structure and mechanics of thermodynamically stable and highly asymmetric soft-hard lamellar structures obtained with optimally designed $PS_1$-$(PI$-$b$-$PS_2)_3$ miktoarm star block copolymers. The mechanical properties of these ductile PS based nanomaterials can be tuned over a broad range by varying the hard layer thickness and keeping the soft layer thickness fixed at 13 nm. The deformation of thin lamellae (< 100 nm) exhibited kinks, pre-damaged/damaged grains, as well as cavitation in soft nano-layers. In contrast, the deformation of thick lamellae (> 100 nm) manifests cavitation in both soft and hard nano-layers. *In situ* tensile-SAXS experiments revealed the evolution of cavities during deformation and confirmed that the damage in such systems reflects both plastic deformation by shear and residual cavities. Although this study was carried out in PS-PI hard-soft systems, the mechanism to obtain asymmetric layered structures should be general and the obtained mechanical properties provide guidance for future designs of nanostructured materials with promising mechanical properties.

# INTRODUCTION

Fabrication of nano-layered materials is an important and increasingly popular topic in nanoscience and nanotechnology.[1-18] The materials are broadly used as coatings, adhesives, membranes, electronic devices and bio-sensors with tunable functionalities, such as solar cell devices (typically multi-nanolayers of conducting materials),[7-9] polyelectrolyte membranes (usually through a layer-by-layer technique),[13-15,18] artificial skins[16,17] and so on. Meanwhile, those materials need to meet various mechanical requirements for different purposes. For example, extensibility and resistance to fracture is critical to develop deformable electronic devices;[8,9] nanomechanical properties and domain features are also sensitive to tune photonic response in polymer multilayers.[10] Recent research has developed to bio-related applications, such as strain sensors for artificial skins.[16] In spite of the broad current and future applications, the fundamental mechanics of lamellar nanomaterials has not yet been well elucidated.[19-32]

From a polymer mechanics perspective, nano-layers of two hard polymers (modulus over ~1 GPa) exhibit high modulus but cannot deform to large strains without fracture; nano-layers of two soft polymers are usually extensible but lack stiffness; a combination of hard-soft polymer nano-layers could synergetically improve the mechanical properties over a broad range.[19-23] Yet the mechanical properties of these structures, in a soft-hard configuration for example, are nearly unexplored and are a prerequisite to expand the applications. Hard-soft block copolymers (BCPs) are good prospects to achieve nano-layered structures by microphase separation.[33-36]

Most current nano-layered materials obtained by traditional block copolymers are however only thermodynamically stable in a nearly symmetric layer thickness, although nanotechnology nowadays requires lamellae with flexibly tunable thickness. In linear block copolymers (such as AB diblocks or ABA triblocks), the available composition range for lamellar structures is usually from 35% to 65% (by volume) in the strong segregation regime.[33] Nano-layered lamellar structures are not stable at higher or lower compositions. How to achieve highly asymmetric lamellae has long been a challenging question in polymer science. Due to this limitation, current understanding on lamellar mechanics is limited in symmetric case, where the inclusion of a high content of soft component improves the toughness and meanwhile results in a considerable loss of modulus and strength. Highly asymmetric hard-soft lamellae, with a high content (above 65%) of a hard component, are expected to manifest a promising balance between toughness and stiffness.[37-43] However, almost nothing has been achieved in this direction. In addition, the mechanics of lamellar structures is highly dependent on the molecular architecture. For hard-soft nano-layers, hard blocks are expected to behave as anchoring points at both ends to provide stiffness, while the middle soft blocks supply extensibility.[37,38,40] The number of hard and soft blocks within one molecule is a key parameter to tune mechanical properties. So, hard-soft diblock copolymers exhibit poor mechanical properties; hard-soft-hard triblock copolymers have improved mechanical properties, but the number of anchoring points per molecules is still low.[35,36]

Guided by self-consistent field theory simulations,[44] we find that the limiting factors described above are addressed with a carefully designed miktoarm architecture $A_1$-(B-$b$-$A_2$)$_n$ (n>2, A is the hard block and B is the soft block, as illustrated in Scheme 1).[37] Driven by the nonlinear molecular architecture at the $A_1B_n$ junctions and the bi-dispersity of A blocks within the molecule, the miktoarm star block copolymer reinforces curvature towards A and stabilizes lamellar structures up to a remarkably high volume fraction $f_A$=80% for n=3 and $\tau = N_{A1}/(N_{A1}+N_{A2})=0.9$, where $N$ is the number of repeating units in each block. The specific realization of this model system was achieved with a $PS_1$-(PI-$b$-$PS_2$)$_3$ system (PS denotes polystyrene and PI poly(isoprene)).[38] Furthermore, blending of $PS_1$-(PI-$b$-$PS_2$)$_3$ with homo-polystyrene (hPS) leads to even more extreme morphologies. The upper boundary of an extremely asymmetric lamellar structure was realized for up to 97 wt% of PS segments in total.[39] Another key aspect of the $PS_1$-(PI-$b$-$PS_2$)$_3$ architecture is that the junctions between PI blocks and the short $PS_2$ tails supply multiple anchoring points in the glassy PS domains so that stress is more uniformly distributed, compared with fewer anchoring points in linear-chain BCP counterparts.

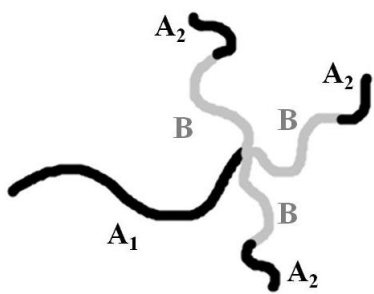

Scheme 1. Schematic illustration of an $A_1$-(B-$b$-$A_2$)$_3$ miktoarm block copolymer.

Based on this molecular design, we would expect the mechanical properties to be tunable over a broad range, combining high extension from the soft rubbery component and high modulus from the hard glassy component. The deformation mechanism is investigated in details using transmission electron microscopy and *in situ* real-time tensile-SAXS synchrotron facility. Notably, although the study is carried out with PS and PI hard-soft systems, the mechanism should be more general and guide the design of other functional materials (such as electric, photonic, permeable) based on lamellar nanostructures with promising mechanical properties. In addition, this study demonstrates the significant advantages of using miktoarm block copolymers as a structuring unit relative to traditional linear block copolymers, resulting in new perspective to design advanced nanomaterials.

## RESULTS AND DISCUSSION

The neat miktoarm block copolymer $PS_1$-(PI-$b$-$PS_2$)$_3$ with 70 wt% PS exhibits an asymmetric lamellar structure. Even more asymmetric lamellae are stable with 90 wt% hPS, which is equivalent to 97 wt% PS in total. For this set of materials, the rubbery PI layer thickness is constant (~13 nm) while the PS layer thickness can be varied from 26 nm to over 300 nm.[39]

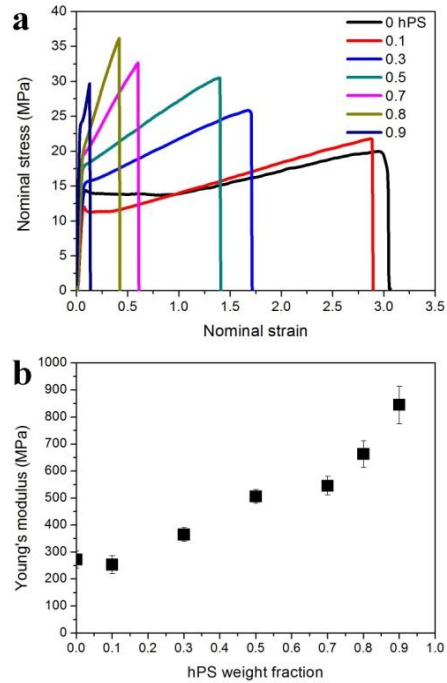

Figure 1. Monotonic tensile testing results for miktoarm block copolymer blends with different weight fractions of hPS: (a) stress-strain curves and (b) Young's moduli.

The monotonic tensile testing results are shown in Figure 1 for miktoarm block copolymer blends with different hPS weight fractions. The neat miktoarm copolymer already exhibits a high stress and strain at break. But due to the significant rubbery content (30 wt% of PI), the modulus, yield stress and nominal strength at break are relatively low as shown in Figure 1. For neat miktoarm and the blend with 10 wt% of hPS, clear plateau regions of nearly constant nominal stress after yield, followed by a mild strain hardening, are observed. The plateau region was accompanied by necking, which indicates that shear yielding is the main deformation mechanism. For blends with hPS content above 30 wt% (rubbery domain volume fraction is now less than 20 wt%), the yield point shifts to larger stresses and pronounced strain hardening is observed after yielding. The plateau region disappears from the stress-strain curves and necking is no longer observed. Remarkably, the material remains ductile with nearly 150% strain at break at an hPS content of 50 wt% (equivalent to 15 wt% rubbery phase). For blends with even more hPS, the nominal strength at break keeps increasing until 80 wt% hPS, while the strain at break continuously decreased. The

Young's moduli of these nano-layered materials increased from ~300 MPa to ~900 MPa (Figure 1b). Compared with 2~3 GPa and ~2% ultimate strain for pure hPS thermoplastics,[19,20] the unusual rubber configuration in these blends is remarkably effective at giving ductility while resulting in only a moderate decrease in modulus relative to the hard PS glassy component. The drop of the modulus will be discussed later.

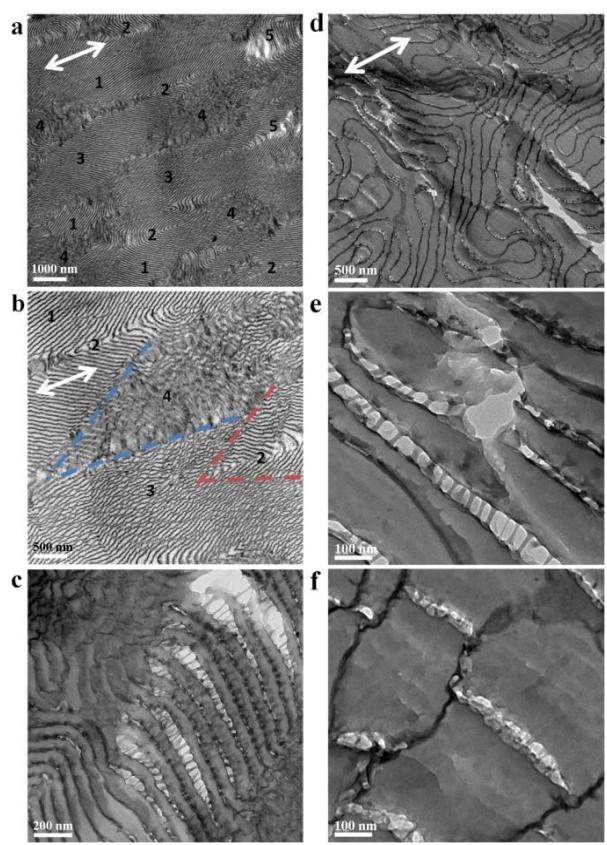

Figure 2. TEM images for the miktoarm block copolymer blended with 50 wt% hPS (a, b, c) and 80 wt% hPS (d, e, f) after stretching until break. The arrow indicates the stretching direction. Image (a) was taken at low magnification showing 5 distinct regions. Image (b) is a magnified image of (a). Image (c) highlights the void formation of a region 5. Image (d) was taken at low magnification. The arrow indicates the stretching direction. Image (e) is a magnified image of (d) highlighting void formation in the rubbery phase. Image (f) is another magnified image of (d) highlighting crazes in the glassy PS domains.

To investigate the fracture mechanism, we studied the structure after tensile deformation (post failure). Figure 2 (a, b, c) show the corresponding TEM images for the sample containing 50 wt% hPS (More evidence was presented as supplementary information: part B). One can clearly indentify five different regions. Region 1 is an intact region where lamellae were well maintained after stretching. Region 2 shows the formation of kinks, which include the original defects and deformed lamellae in the tensile stress direction.[45-47] It is notable that the lamellae near the kinks exhibit minor damage. In region 3, we observe weakly deformed lamellae with undulating interfaces which manifest a stress-induced instability. In these regions, the lamellae are relatively well aligned with the tensile stress direction. We denote region 3 as a "pre-damage grain." Region 4, neighboring region 3, contains lamellae that were more severely damaged resulting in homogenized grains with greatly reduced orientational order. The originally sharp PS/PI interfaces are smeared and distorted within such grains. We denote region 4 as a "damage grain." V-shaped boundaries between the regions 3 and 4 can be clearly identified (Figure 2b). The deformation behavior of regions 3 and 4 are very similar to the shear deformation of semi-crystalline polymers under tensile stress.[32]

Finally, we consider regions (5) where the lamellae are oriented perpendicular to the tensile stress direction. As shown in Figure 2c, the PS glassy layers remained largely intact while voids grew within the PI rubbery domains. The PI domains form fibrils between neighboring PS layers and can extend to several times their original thickness before breaking. The original PI layer thickness is 13 nm. From the TEM images, we observe that the PI fibrils break when the extension is ~100 nm, which indicates an extension ratio of approximately 8. Further tensile deformation evidently leads to fibril breakdown and coalescence of neighboring voids into cracks. It appears, therefore, that the fracture of the material is initiated from region 5. Local shear deformation is still a dominant factor under tensile deformation, in spite of the fact that brittle PS constitutes the majority phase (85 wt% PS in total).

When the hPS fraction was increased to 80 wt% (94 wt% PS in total), we found that the fracture mechanism changed considerably. Figure 2 (d, e, f) show the corresponding TEM images for the 80 wt% hPS sample after monotonic tensile testing to break. The average PS layer thickness is now approximately 170 nm while the PI layer thickness is 13 nm. In Figure 2d, we observed significant void formation in the rubbery layers; there are also clear indications of voids in the glassy PS domains. We can also observe a crack tip (right bottom) expanding into both PS and PI regions. In Figure 2e, the rubbery fibrils seem to remain intact at small extension, and fail when the thickness of the fibrillar region was larger than ~100 nm. Crazes, consisting of PS fibrils and voids between two interfaces, can be identified in Figure 2f. The size of the crazes is about ~50 nm. The craze phenomenon of the PS nano-layers indicates a similar mechanical response than that of the bulky PS. Notably, the interfaces between glassy and rubbery domains remain almost intact, a significantly different behavior from that observed for the blends with lower (50 wt%) hPS content.

From the above comparison, we conclude that the tensile deformation of the blends with relatively low hPS content is characterized by two features: 1) the extension of the rubbery nanolayers and 2) the significant deformation of the interfacial regions, which leads to the formation of pre-damage and damage grains. Both factors dissipate energy and assist the release of local stress during stretching, which cause irreversible damage in the material. In contrast, for the blends with high hPS content (above 80 wt%), the interface regions remain largely intact, while crazing is observed in the relatively thick PS domains (> 100 nm). Although PS crazes can still sustain a load at small strains, the nano-fibrils will break at larger deformation and result in catastrophic coalescence of voids into growing cracks (Figure 2d). The void formation in both PS and PI layers acts potentially as 3-dimensional channels for crack propagation. So, in the high hPS content blends, the observed toughness mainly relies on the extension of the PI rubbery lamellae (8~10 times of extension).

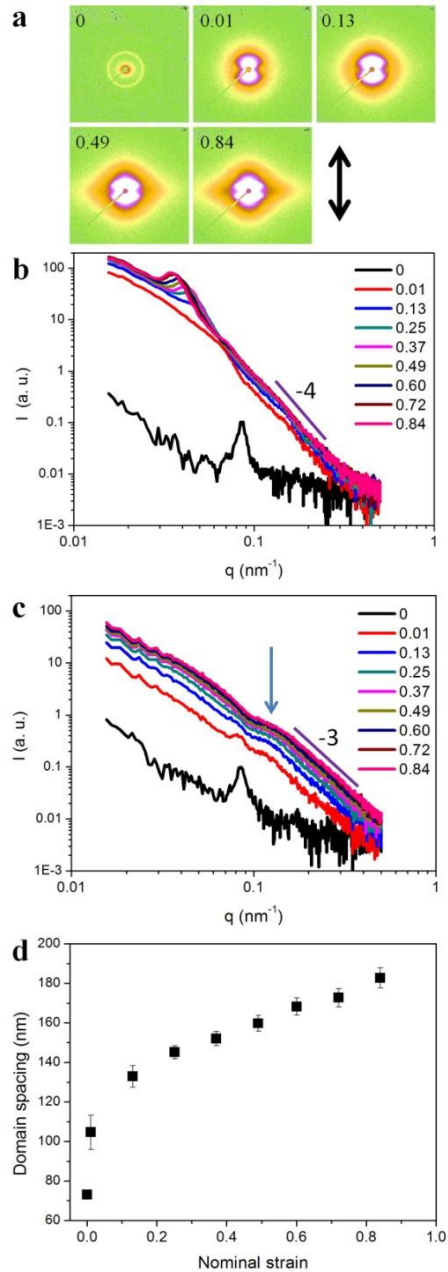

Figure 3. Tensile-SAXS experiments for the miktoarm block copolymer blended with 50 wt% hPS. The 2-D SAXS patterns are shown in image (a) at different nominal strains. The arrow indicates the tensile direction. Images (b) and (c) are scattering curves in the tensile direction and transverse directions, respectively. The arrow in image (c) marks the scattering peak associated with correlation of voids. Image (d) shows the domain spacing variation with nominal strain. The domain spacing was calculated from the primary peak positions in the SAXS curves shown in image (b).

To further investigate the void formation during stretching, we carried out *in situ* real-time tensile-SAXS experiments, which provide instantaneous scattering patterns under deformation. Figure 3 shows the scattering patterns for miktoarm/hPS (50 wt%) blends at different strains. The original sample showed isotropic scattering rings, indicating lamellar structures with random orientation (Figure 3a). Upon stretching, the intensity grew strongly in the tensile direction and exhibited "butterfly" patterns. The anisotropic scattering patterns are the result of increased correlation and contrast in the tensile direction, a sign of void formation. The 2-D SAXS patterns were plotted as 1-D curves in the tensile and transverse directions in Figures 3b and c, respectively. The intensity in the tensile direction is observed to be about 10 times stronger than that in the transverse direction near $q=0.02$ nm$^{-1}$. In Figure 3b, the primary peak position is seen to shift toward smaller scattering wavenumber q with growing intensity; the slope in the high q region followed a power law index -4, indicative of scattering from sharp interfaces. We postulate the primary peak is ascribed to the increasing domain spacing associated with substantial extension of PI layers oriented perpendicular to the tensile direction (filled with PI fibrils and voids). The initial domain spacing is 73 nm, where the thicknesses of the PS and PI layers are 60 and 13 nm, respectively. As shown in Figure 3d, the domain spacing increased significantly upon stretching, with the increment slowing after yielding, and finally reaching 182 nm before break. We infer that the PI fibrils can extend to 122 nm until fracture of the sample. These results are consistent with the TEM images (Figure 2) showing void formation in the rubbery phase and PI fibrils stretched between glassy domains with extension ratios of 8~10. In Figure 3c, we observe only weak, broad peaks in the transverse scattering, which suggests low contrast or smeared lamellar order; the slope in the high q region shows a power law index -3, which is characteristic of scattering from diffuse/fractal interfaces.

The scattering patterns generally exhibited similar features for other blends at lower or higher hPS contents (supplementary information: part C). Notably, for a range of hPS compositions, a weak, broad peak always appeared at ~0.12 nm$^{-1}$ in the transverse direction during stretching (Figure 3c), which corresponds to a correlation length of about 50 nm. Since TEM analysis reveals that the domain features are quite different in high and low hPS content miktoarm/hPS blends and the only common feature arises from the void formation, we postulate that the transverse scattering peak reflects the correlations between neighboring voids (not the form factor of individual voids, since the peak position did not move with strain). The correlation length is ~50 nm, which is consistent with TEM images.

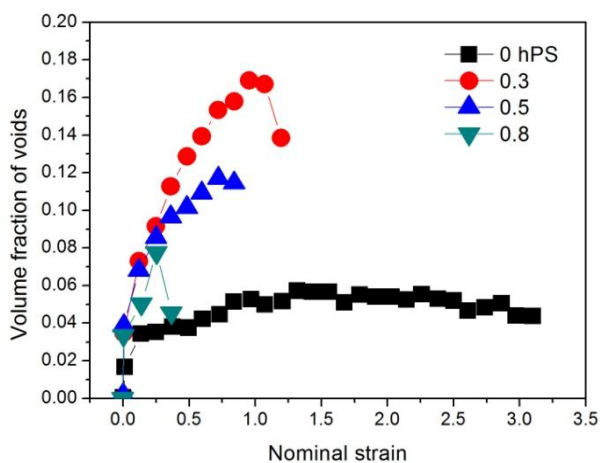

Figure 4. The volume fraction of voids during monotonic tensile stretching for miktoarm/hPS blends at different hPS weight fractions.

Using the well established three-phase model,[48,49] we can quantitatively calculate the volume fraction of voids during tensile stretching with the full 2-D scattering data (Figure 4). The neat miktoarm block copolymer had low void fraction around 5% in a large range of strains. The void content increased significantly when hPS was blended with the miktoarm copolymer. For 30 wt% hPS blends, the fraction could reach 17% before rupture. For blends with even higher hPS content, the void content followed a similar increasing tendency with strain but the material broke at smaller strains. One common feature can be clearly identified for all blends: the void formation follows a similar strain-dependence before yielding, which indicates that the void formation mechanism is similar for all blends at small strains. From Figure 3d, we found that the domain spacing increased quickly before yielding. So, cavitation in the rubbery nano-layers was the major contribution to the total void fraction for all miktoarm/hPS blends.

To gain further insight into the deformation mechanism, we carried out SAXS experiments during step-cycle tensile testing. For the neat miktoarm copolymer, the strain increment in each cycle was 0.5; that is, the sample was stretched to a strain of 0.5 and then the force was reduced to 0 in the first cycle; in the second cycle, the sample was stretched to a strain of 1.0 and then the force was reduced to 0; the maximum strain is 1.5 in the third cycle; and the process was repeated until the rupture of the specimen. For blends with 50 wt% and 80 wt% hPS, the strain increment was 0.12 for each cycle. The volume fraction of voids was calculated throughout the repeated load-unload processes and the results are shown in Figure 5.

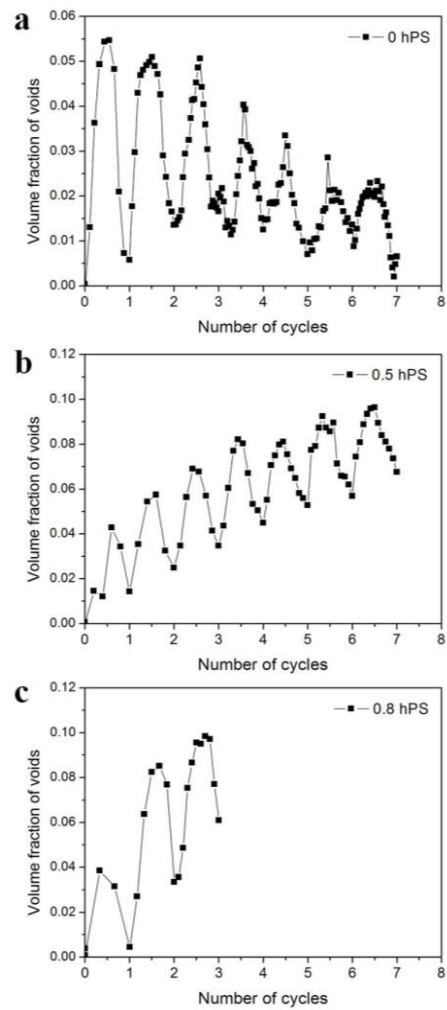

Figure 5. The void content during step-cycle tensile tests for miktoarm/hPS blends at different hPS weight fractions. For neat miktoarm copolymer (a), the strain increment

was 0.5 for each cycle; for blends with 50 wt% (b) and 80 wt% hPS (c), the strain increment was 0.12 for each cycle.

In the neat miktoarm copolymer (Figure 5a), the voids were significantly initiated in the first loading cycle, and then almost fully disappeared after the first unloading. In the next multiple loading-unloading cycles, the maximum volume fraction of voids kept decreasing. In the seventh cycle, the maximum void content was about 2% and could be fully recovered after unloading. From the lamellae "melting" under deformation (damage grains in Figure 2), we infer that the local stress inhomogeneity was significantly released by breaking the lamellar domains therefore the voids cannot be initiated in the already homogenized grains. When the hPS content is increased, it becomes difficult to reversibly deform hard PS layers and the readily formed voids do not close upon unloading. Void formation in both PS and PI domains becomes prominent as the amount of hPS increases. For blends with 50 wt% (Figure 5b) and 80 wt% hPS (Figure 5c), we found that although the voids could almost be fully recovered in the first loading-unloading cycle (10% strain), the void content kept increasing with strain for each subsequent tensile cycle and only part of the voids closed after each unloading. By further analysis of the SAXS data for each cycle (Supplementary information: part D), we found evidence of a strong effect of the mechanical history on the structure along the tensile direction, while the mechanical history had a much weaker effect on the transverse structure. Notably, the void correlation peak at 0.12 nm$^{-1}$ is clearly visible upon loading but is attenuated after unloading. Here we conclude two points: the non-recovery of miktoarm/hPS blends reflects both plastic deformation and residual voids; voids can be partially kept open under hydrostatic stress in the tensile direction, while shear deformation dominates in the transverse direction. The latter point is also supported by TEM images (Figure 2).

As stated above, PS layers tend to be destroyed rather than forming crazes when the PS layers are thin; crazing in PS domains becomes prominent when the thickness of the PS layers is large enough. A key question remains: what is the critical thickness separating the two regimes? In different, but related systems, researchers have proposed that glassy lamellae should be thin enough to avoid brittle fracture by crazing.[41-43] They suggested a critical ligament thickness around 20 nm, which is approximately the craze fibril thickness in neat hPS. We do not find evidence supporting this 20 nm value in our systems. For the asymmetric miktoarm block copolymer blended with 50 wt% hPS, the thickness of the PS layers is 60 nm, which is a value three times higher than the proposed critical thickness, yet the material exhibits substantial extension with no PS crazing detected. In contrast, PS crazing was clearly observed in blends containing 80 wt% hPS, with PS domain thickness of ~170 nm. Based on these observations, we propose that the PS critical thickness in asymmetric lamellar systems defining a ductile to brittle transition is roughly about 100 nm.

Combining all the above results and analysis, we can classify five categories of lamellar deformation under tensile stretching:

Ⅰ. *Elastic deformation*. Elastic deformation is only maintained at small strains before any irreversible damage occurs.

Ⅱ. *Orientational reorganization*. Under stretching, the domains will adapt to the local shear or compression/extension forces by changing orientation. The observed kinks are typical features associated with orientational reorganization of lamellar structures.[45-47]

Ⅲ. *Mechanical homogenization*. If the chain density at the interface is low and a lamellar region is oriented poorly, short PS blocks can be pulled out from their domains[50,51] and PS layers can be fragmented, leading to the "melting" of a lamellar structure into homogenized state. In this study, regions 3 and 4 in Figure 2a show evidence of this kind of deformation.

Ⅳ. *Void formation in rubbery domains*. Since the rubbery phase has a low modulus and is confined between hard PS layers, it is easy to deform it under stress. A void can "nucleate" where the local hydrostatic tensile stress is sufficiently high. Once nanovoids are formed, the rubbery phase can continue sustain load until the fibrils break.[19,48,49,52] An open question left here is: where does the nucleation of voids primarily occur? A recent simulation work demonstrates that the primary positions of voids should be at the interface between soft and hard domains,[53] which still need experimental confirmation.

Ⅴ. *Void formation in the glassy domains*. Because of the high modulus of the glassy domains, a high local stress (above the crazing stress of PS) is needed to form crazes in the glassy domains.[19] For small PS domain thicknesses, the local stress can be relieved by "melting" of PS layers; but crazes in the glassy domains dominate the fracture mechanism when the PS domains are thick enough. From our study, the critical thickness is in the vicinity of 100 nm.

As a final note, compared with neat PS, the asymmetric PS-PI lamellae exhibit a smaller modulus even with 3 wt% soft PI component (90 wt% hPS in this case). The relatively small modulus, we postulate, is due to the presence of voids initiated at small strains (<1%, see Figure 3) in the soft PI nanolayers deformed in tension. Such early damage gives an apparent modulus smaller than neat PS.

## CONCLUSION

In conclusion, we demonstrate that well designed $PS_1$-$(PI$-$b$-$PS_2)_3$ miktoarm star block copolymers can be remarkably effective at toughening PS homopolymer, even when the blend has a very low rubber concentration (3 to 15 wt% PI). The ductility arises from highly asymmetric lamellar morphologies perfused with continuous rubbery PI sheets and represents a new paradigm for creating tough, hard, and strong material from intrinsically brittle glassy polymers. The toughening mechanism for these novel thermoplastics primarily relies on the extensive deformation of the rubbery domains and mechanical homogenization. Deformation of the rubbery layers leads to significant void formation, but the rubbery fibrils sustain significant stress until break. Mechanical homogenization of the glassy and rubbery nano-layers leads to the formation of pre-damage and damage grains, which dissipate energy and prevent the formation of crazes in the glassy domains. However, when the PS nano-layers are sufficiently thick (> 100 nm), crazing in PS domains becomes dominant and leads to catastrophic fracture of the material. The specific hard-soft asymmetric nano-layered polymer systems in this study present a powerful protocol for revealing the structure-mechanics relationship. The mechanism should be ubiquitous to direct other nanomaterials with desired mechanical properties.

## EXPERIMENTAL SECTIION

**Materials.** The $PS_1$-$(PI$-$b$-$PS_2)_3$ miktoarm star block copolymers were synthesized by anionic polymerization, which was reported elsewhere.[38,39] The long $PS_1$ block has a molecular weight of ~ 80 kg/mol and the short $PS_2$ blocks have a molecular weight of ~ 10 kg/mol. The length of the middle PI block was adjusted to control and vary the soft/hard volume fraction. In this study, for the highly asymmetric lamellar thermoplastics, we used a miktoarm copolymer with an overall PS volume fraction of 0.67 (70 wt% PS) and hPS with a molecular weight of 81.7 kg/mol. The detailed molecular characterization is provided in supplemental information (Part A).

**Sample preparation.** Polymer blends with various hPS fractions were prepared in toluene and stirred at room temperature overnight. A small amount of BHT (< 0.5%) was added to prevent the oxidation of PI blocks. The mixtures were poured on Teflon films and further dried at a fume hood for one week to evaporate the solvent. The thermal annealing was carried out at 150 °C for 48 h in a high vacuum chamber ($10^{-8}$ mbar).

**Mechanical tensile testing.** The annealed films were punched with a metal die to produce dog-bone shaped specimens. Each specimen had 7 mm gauge length and a cross-section of 2 mm wide by ~ 0.5 mm thick. Monotonic and step-cycle tensile tests were carried out with 5 specimens on a home-made tensile apparatus. The load cell had capacity of 40 N. The bottom clamp was stationary while the upper one was movable. The crosshead speed was maintained at 5 mm/min for all the specimens, producing an initial strain rate of 0.012 $s^{-1}$.

**Transmission electron microscopy (TEM).** After the tensile testing, the broken specimens were collected and stained in osmium tetroxide aqueous solution (2 wt%) for one week. The $OsO_4$ selectively stains PI (due to cross-linking of the vinyl bonds with the oxygen atoms of the $OsO_4$ molecule), so the structures were fixed and prevented relaxation or mechanical deformation during sectioning by microtome. The stained specimens were subsequently cut on an ultra-microtome machine at room temperature to produce ultrathin (~100 nm) slices. The thin slices were collected on copper grids for TEM observation. The PI domains appear dark in the TEM images.

**Real-time tensile-SAXS experiments.** The synchrotron small angle x-ray scattering (SAXS) experiments were carried out using the Advanced Photon Source beamline 5-ID at Argonne National Laboratory. The wavelength of the source was 0.124 nm. The sample to detector distance was set to 7496 mm so that the lowest wave vector was $0.0155 nm^{-1}$, which is equivalent to 400 nm. The beamline was adapted to an Instron 850 tensile tester. The load cell capacity was 200 N. Both upper and bottom clamps were movable to create symmetric tensile stretching. The strain rate was kept at $0.012 s^{-1}$ for all tests.

**Corresponding Authors**
*E-mail: ghf@mrl.ucsb.edu (GHF); aavger@cc.uoi.gr (AA)
**Notes**
The authors declare no competing financial interest.

**Acknowledgements**
This research was supported by the Institute for Collaborative Biotechnologies through grant W911NF-09-0001 from the U.S. Army Research Office. The content of the information does not necessarily reflect the position or the policy of the Government, and no official endorsement should be inferred. Extensive use was made of the MRL Shared Experimental Facilities supported by the MRSEC Program of the NSF under Award No. DMR 1121053; a member of the NSF-funded Materials Research Facilities Network (www.mrfn.org). This work was performed at the DuPont-Northwestern-Dow Collaborative Access Team (DND-CAT) located at Sector 5 of the Advanced Photon Source (APS). DND-CAT is supported by E.I. DuPont de Nemours & Co., The Dow Chemical Company and Northwestern University. Use of the APS, an Office of Science User Facility operated for the U.S. Department of Energy (DOE) Office of Science by Argonne National Laboratory, was supported by the U.S. DOE under Contract No. DE-AC02-06CH11357.

**Supporting Information Available**

Part A: Samples used and molecular characterization; Part B: TEM image of miktoarm/hPS (50 wt%) blends before/after deformation; Part C: Monotonic tensile-SAXS data for miktoarm/hPS (80 wt%) blends; Part D: Step-cycle tensile-SAXS data for miktoarm/hPS (50 wt%) blends.

ToC

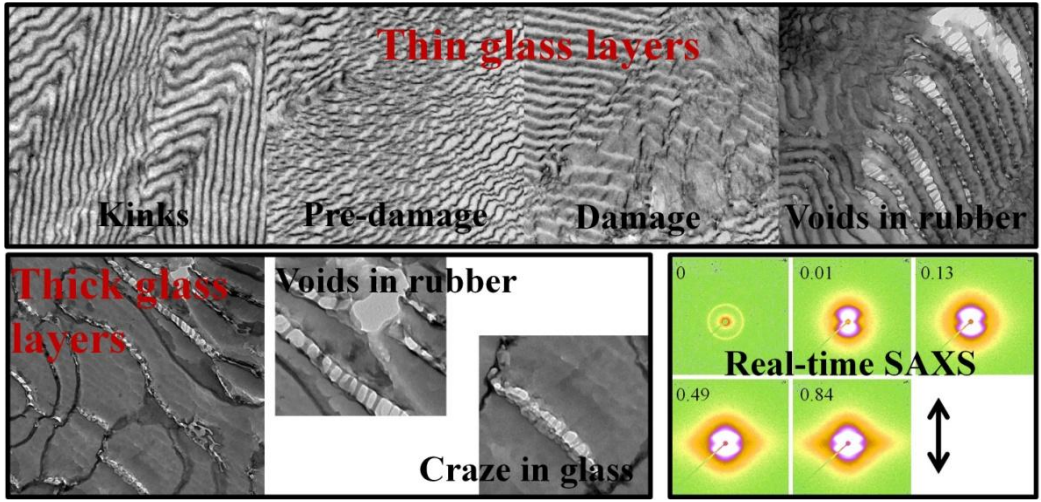